# Study on the spectral reconstruction of typical surface types based on spectral library and principal component analysis

Weizhen Hou[a,b], Yilan Mao[c], Chi Xu[c], Zhengqiang Li[*,a,b], Donghui Li[a], Yan Ma[a], Hua Xu[a]
[a]State Environmental Protection Key Laboratory of Satellite Remote Sensing, Institute of Remote Sensing and Digital Earth, Chinese Academy of Sciences, Beijing, China 100101; [b]State Key Laboratory of Remote Sensing Science, Institute of Remote Sensing and Digital Earth, Chinese Academy of Sciences, Beijing, China 100101; [c]Beijing Institute of Spacecraft Systems Engineering, Beijing, China 100094

## ABSTRACT

To meet the demanding of spectral reconstruction in the visible and near-infrared wavelength, the spectral reconstruction method for typical surface types is discussed based on the USGS/ASTER spectral library and principal component analysis (PCA). A new spectral reconstructed model is proposed by the information of several typical bands instead of all of the wavelength bands, and a linear combination spectral reconstruction model is also discussed. By selecting 4 typical spectral datasets including green vegetation, bare soil, rangeland and concrete in the spectral range of 400–900 nm, the PCA results show that 6 principal components could characterized the spectral dataset, and the relative reconstructed errors are smaller than 2%. If only 6–7 selected typical bands are employed to spectral reconstruction for all the surface reflectance in 400–900 nm, except that the reconstructed error of green vegetation is about 3.3%, the relative errors of other 3 datasets are all smaller than 1.6%. The correlation coefficients of those 4 datasets are all larger than 0.99, which can effectively satisfy the needs of spectral reconstruction. In addition, based on the spectral library and the linear combination model of 4 common used bands of satellite remote sensing such as 490, 555, 670 and 865 nm, the reconstructed errors are smaller than 8.5% in high reflectance region and smaller than 1.5% in low reflectance region respectively, which basically meet the needs of spectral reconstruction. This study can provide a reference value for the surface reflectance processing and spectral reconstruction in satellite remote sensing research.

## 1. INTRODUCTION

The spectral library is a collection of various types of surface reflectance spectral dataset measured by the hyperspectral imaging spectrometer or field spectrometer under certain conditions, which can accurately interpret remote sensing image information, quickly realize the matching of unknown features, and improve the remote sensing classification and recognition[1-3]. The surface spectrum is an important part of the basic research of remote sensing, and the corresponding spectral library have been established both domestically and internationally[4-8].

Principal component analysis (PCA) has been widely used in the study of spectral compression and reconstruction, in which, a spectral data set can be decomposed into the linear combinations of several main spectral components[9-12]. Previous studies have shown that in the visible range of 400–700 nm, the first six principal components (PCs) can meet the needs of spectral reconstruction[13-14]; while for the wavelength range of 400–2500 nm, 12–20 PCs are needed to reconstruct the vegetated surface spectra[15].

In this paper, based on the surface reflectance dataset in the spectral library and PCA method, the typical spectral reconstruction method in the range of 400–900 nm is studied, and the spectral reconstruction results are analyzed and evaluated in this paper, which can provide the support for subsequent hyperspectral and multispectral remote sensing applications.

## 2. DATA

At present, the most commonly used surface spectral library of includes the USGS spectral library and the ASTER spectral library. The USGS spectral library (http://speclab.cr.usgs.gov/spectral.lib06/) is established by the United State Geological Survey (USGS) Spectroscopy Laboratory in the wavelength range of 0.2–3.0 μm, and the spectra can be mainly divided in to six categories, including minerals, mixtures, coasting, volatiles, man-made materials and vegetations[4-6]. The ASTER spectral library (http://speclib.jpl.nasa.gov) is part of Advanced Space-borne Thermal Emission and Reflection Radiometer (ASTER) research program by NASA, which mainly includes the spectra of minerals, rocks, soils, man-made materials, water, ice and snow, many of which can cover the wavelength from 0.4 to 14 μm[7]. By resampling the spectra data from these two typical spectral libraries, the spectral datasets with different spectral resolutions can be obtained for the hyperspectral or multispectral remote sensing.

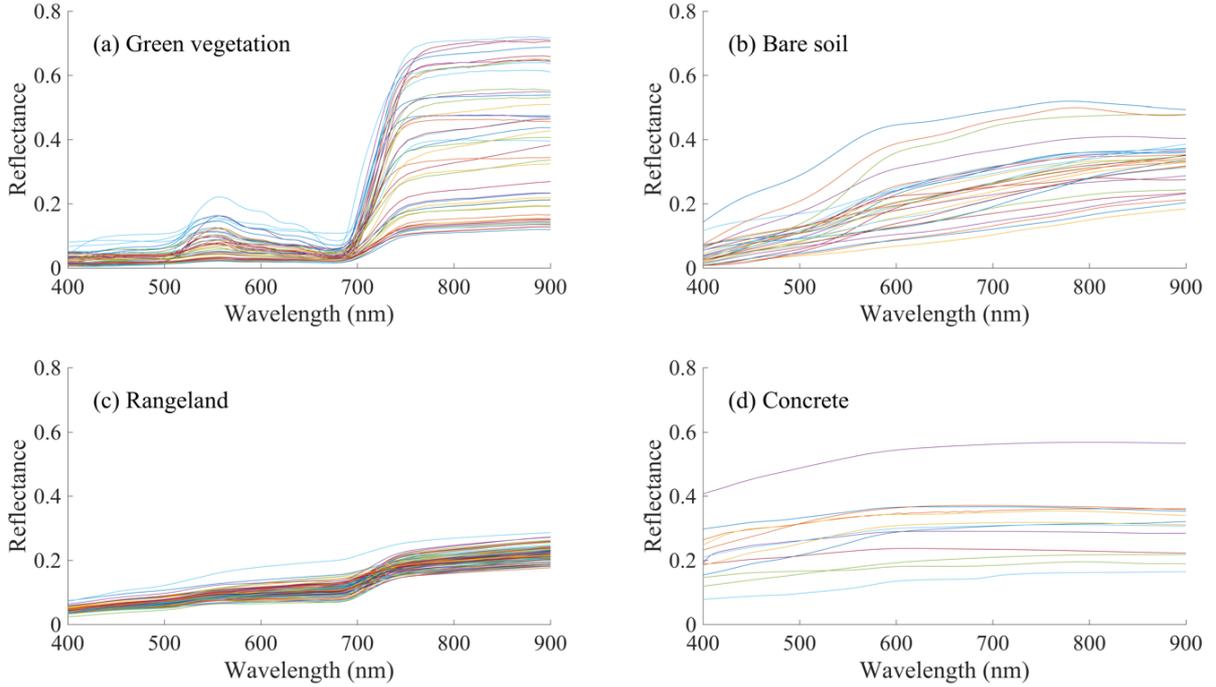

Figure 1. The spectral datasets of 4 typical surface types from USGS/ASTER spectral library. (a) green vegetation, (b) bare soil, (c) rangeland, (d) concrete.

As shown in Figure 1, we consider four typical surface spectral dataset, including green vegetation, bare soil, rangeland and concrete, with the re-sampling spectral resolution of 1 nm in the visible and near-infrared range of 400–900 nm in the study, and the spectral curves in the dataset are 50, 30, 84 and 13, respectively. Among them, the spectral datasets of green vegetation and rangeland are selected from the USGS spectral library, in which the rangeland represents a mixed surface type composed of grassland, shrub, woodland, wetland and desert. Besides, the bare soil spectral dataset is selected from the ASTER spectral library and the spectral smoothing has been further carried out, while the concrete spectral dataset is selected from both spectral libraries to represent typical urban bright surface types.

## 3. METHODOLOGY

### 3.1 Principal component analysis of spectral dataset

For the principal component analysis method with the same surface type in the range of 400–900 nm, suppose we have a sample set $X$, which have $n$ samples, and each sample has $d$ dimensions, that is

$$X = \{X_1, \cdots, X_n\}, X_i = (x_{i,1}, \cdots, x_{i,d})^T \in R^d, (i = 1, \cdots, n) \tag{1}$$

where the superscript $T$ represents the transpose operation. These samples can be expressed in a matrix form as

$$\mathbf{S} = \begin{bmatrix} X_1^T \\ \vdots \\ X_n^T \end{bmatrix} = \begin{bmatrix} x_{1,1} & \cdots & x_{1,d} \\ \vdots & \ddots & \vdots \\ x_{n,1} & \cdots & x_{n,d} \end{bmatrix} \tag{2}$$

and the mean of the vectors can be written as

$$\bar{X} = \left[\frac{1}{n}\sum_{i=1}^{n} x_{i,1}, \cdots, \frac{1}{n}\sum_{i=1}^{n} x_{i,d}\right]^T = [\bar{x}_1, \cdots, \bar{x}_d]^T \tag{3}$$

Thus, the covariance matrix $\mathbf{C} \in R^{d \times d}$ can be represented as

$$Cov(x_j, x_l) = C_{j,l} = \frac{1}{n-1}\sum_{i=1}^{n}(x_{i,j} - \bar{x}_j)(x_{i,l} - \bar{x}_l), \ (j = 1, \cdots, n; l = 1, \cdots, n) \tag{4}$$

For the principal component analysis (PCA), we diagonalize the covariance matrix $\mathbf{C}$, here the matrix $\mathbf{C}$ is a symmetric matrix, and what we need is to find an orthogonal matrix $\mathbf{P}$ to satisfy the condition that:

$$\mathbf{P}^T \mathbf{C} \mathbf{P} = \mathbf{\Lambda} = diag(\lambda_1, \cdots, \lambda_d), \lambda_1 \geq \cdots \geq \lambda_d \tag{5}$$

Thus, we can eigen-decompose the matrix $\mathbf{C}$ to get the diagonal eigenvalues matrix $\mathbf{\Lambda}$, and orthogonalize the corresponding eigenvectors to get the matrix $\mathbf{P} = [\mathbf{p}_1, \cdots, \mathbf{p}_d]$, the orthogonalized eigenvector $\mathbf{p}_i$ can be called the principal component (PC)[9-12]. Correspondingly, the cumulative contribution rate of former m principal components can be written as

$$v_m = \sum_{i=1}^{m} \lambda_i / \sum_{i=1}^{d} \lambda_i \tag{6}$$

### 3.2 Spectral reconstruction model based on all bands

Following the procedure of PCA, the surface reflectance at each wavelength band for a given sample could be represented and reconstructed by the PCs as

$$\mathbf{r} = \tilde{\mathbf{r}} + \mathbf{\varepsilon}_r = \mathbf{Pw} + \mathbf{\varepsilon}_r \tag{7}$$

where $\mathbf{r}$ is the vector of surface reflectance spectrum, $\tilde{\mathbf{r}}$ is constructed vector by $m$ PCs, $\mathbf{P}$ is the matrix that constituted by each used PC as the column vector, $\mathbf{w}$ represents the weighting coefficients vector corresponding to the PCs, and $\mathbf{\varepsilon}_r$ means the vector of reconstructed errors[13-14].

If we only consider the first $m$ PCs, Eq. (7) can be written as the vector and matrix form

$$\begin{bmatrix} r_{\lambda_1} \\ \vdots \\ r_{\lambda_d} \end{bmatrix} = \begin{bmatrix} \tilde{r}_{\lambda_1} \\ \vdots \\ \tilde{r}_{\lambda_d} \end{bmatrix} + \mathbf{\varepsilon}_r = \begin{bmatrix} P_{1,1} & \cdots & P_{1,m} \\ \vdots & \ddots & \vdots \\ P_{d,1} & \cdots & P_{d,m} \end{bmatrix} \begin{bmatrix} w_1 \\ \vdots \\ w_m \end{bmatrix} + \mathbf{\varepsilon}_r, (m \leq d) \tag{8}$$

here $r_{\lambda_i}$ represents the surface reflectance at the wavelength $\lambda_i$, the subscript $i$ represents index for the $i^{th}$ wavelength with $i = 1, \cdots, d$ and $d = 501$. Since $\mathbf{P}$ is an orthogonal matrix by the column vector, it satisfies the condition that $\mathbf{P}^T \mathbf{P} = \mathbf{E}$, here $\mathbf{E}$ is an identity matrix, also called unit matrix. Therefore, $\mathbf{w}$ can be calculated $\mathbf{w}$ by

$$\mathbf{w} = \mathbf{P}^T \tilde{\mathbf{r}} \approx \mathbf{P}^T \mathbf{r} \tag{9}$$

Obviously, if $m$ is equal to $d$, the corresponding spectral reconstruction error is zero.

### 3.3 Spectral reconstruction model based on selected bands

Based on Eq. (8) and the PC results of each wavelength band in the range of 400–900 nm, if the surface reflectance information of several typical bands commonly used by satellite remote sensing is used to reconstruct the spectra from 400 nm to 900 nm, the corresponding weighting coefficient vector $\mathbf{w}$ must be obtained first. Corresponding to Eq. (7) and Eq. (8), we have

$$\mathbf{\rho} = \mathbf{P}_\rho \mathbf{w} + \mathbf{\varepsilon}_\rho \tag{10}$$

and

$$\begin{bmatrix} \rho_{d_1} \\ \vdots \\ \rho_{d_k} \end{bmatrix} = \begin{bmatrix} P_{d_1,1} & \cdots & P_{d_1,m} \\ \vdots & \ddots & \vdots \\ P_{d_k,1} & \cdots & P_{d_k,m} \end{bmatrix} \begin{bmatrix} w_1 \\ \vdots \\ w_m \end{bmatrix} + \boldsymbol{\varepsilon}_\rho \tag{11}$$

where $\boldsymbol{\rho}$ represents the vector consisting of the elements for $k$ selected wavelength bands ($k \ll d$) with the subscripts sequentially marked as $d_1, d_2, \cdots, d_k$, $\mathbf{P}_\rho$ is the corresponding PC submatrix, and $\boldsymbol{\varepsilon}_\rho$ is the error vector. Thus, we can use the selected vector $\boldsymbol{\rho}$ to approximately determine the result of $\mathbf{w}$, and then to reconstruct the whole surface reflectance spectra. Following the relationship of Eq. (10), the weighting coefficients vector $\mathbf{w}$ could be approximately solved as a least square solution

$$\mathbf{w} \approx \mathbf{P}_\rho^+ \mathbf{r} \tag{12}$$

where $\mathbf{P}_\rho^+$ means a generalized inverse of matrix $\mathbf{P}_\rho$, and can be calculated by

$$\mathbf{P}_\rho^+ = \left(\mathbf{P}_\rho^* \mathbf{P}_\rho\right)^{-1} \mathbf{P}_\rho^* \tag{13}$$

and $\mathbf{P}_\rho^*$ means the conjugate transpose of $\mathbf{P}_\rho$.

### 3.4 Reconstruction model based on a linear combination

If the surface reflectance $\rho_b$ (subscript marked as b) in the selected $k$ wavelength bands is still unknown, a linear combination model of surface reflectance is necessary

$$\rho_b = \sum_{i=1}^{\tilde{k}} a_i \rho_i \tag{14}$$

where $\rho_i$ means the known surface reflectance from the spectral dataset with the band number $i = 1, \cdots, \tilde{k}$ and $\tilde{k} < k$, the vector $\boldsymbol{a} = [a_1, \cdots, a_{\tilde{k}}]^T$ means the corresponding weighting coefficients. Based on the surface spectral information from the spectral library, we can have the relationship that

$$\begin{bmatrix} \rho_{b,1} \\ \vdots \\ \rho_{b,n} \end{bmatrix} = \begin{bmatrix} \rho_{1,1} & \cdots & \rho_{\tilde{k},1} \\ \vdots & \ddots & \vdots \\ \rho_{1,n} & \cdots & \rho_{\tilde{k},n} \end{bmatrix} \begin{bmatrix} a_1 \\ \vdots \\ a_{\tilde{k}} \end{bmatrix} + \boldsymbol{\varepsilon}_\rho \tag{15}$$

where $[\rho_{b,1}, \cdots, \rho_{b,n}]^T$ means the vector contains $n$ elements at the No. b band from spectral dataset, $[\rho_{i,1}, \cdots, \rho_{i,n}]^T$ means the column vector at No. $i$ band with $i = 1, \cdots, \tilde{k}$, and the column vectors further form the matrix in the right side of Eq. (15). Thus, the coefficient vector of the linear combination can be obtained as

$$\begin{bmatrix} a_1 \\ \vdots \\ a_{\tilde{k}} \end{bmatrix} \approx \begin{bmatrix} \rho_{1,1} & \cdots & \rho_{\tilde{k},1} \\ \vdots & \ddots & \vdots \\ \rho_{1,n} & \cdots & \rho_{\tilde{k},n} \end{bmatrix}^+ \begin{bmatrix} \rho_{b,1} \\ \vdots \\ \rho_{b,n} \end{bmatrix} \tag{16}$$

where the operator $[\ ]^+$ means the corresponding generalized inverse of matrix, just same as the definition in Eq. (13).

## 4. RESULTS

### 4.1 Spectral reconstruction results with all bands

Taking the two spectra datasets of green vegetation and bare soil as an example, Figure 1 shows the corresponding results of PCA, while the other two typical datasets of rangeland and concrete also have similar results and are not shown here. From Figure 2 (a–b), the variance contribution rate of the first PC is largest, which has exceeded 95%, and the contribution of other PC decrease in turn as the order of the PC increases. The contribution of the 6th PC is only about 0.02%, while the total variance contribution rate of the first six PCs has been up to 99.98%. Besides, Figure 2 (c–d) illustrate the spectra of the first six PCs, and thus each spectral curve in the spectral dataset could be reconstructed form these six PCs in combination with the weighting coefficients. In addition, Figure 2 (e–f) show that the reconstruction error decreases as the number of PCs increases. When the first six PCs are used for spectral reconstruction of green vegetation dataset, the mean relative error is less than 2%. While for bare soil dataset, the mean reconstruction relative error has been less than 1% just with 4 PCs. Here, the mean relative error corresponds to the mean value of the relative errors obtained by the leave-one-way cross validation for spectral reconstruction.

Furthermore, Figure 3 shows the scatterplots of the reconstructed reflectance with 1 PC to 6 PCs versus the true reflectance over 501 spectral wavelengths in the range of 400–900 nm for the green vegetation and bare soil surface dataset respectively, and the correlation coefficients ($R^2$) are all larger than 0.999. Based on the above analysis, the first six PCs can fully characterize the spectral datasets in the range of 400–900 nm.

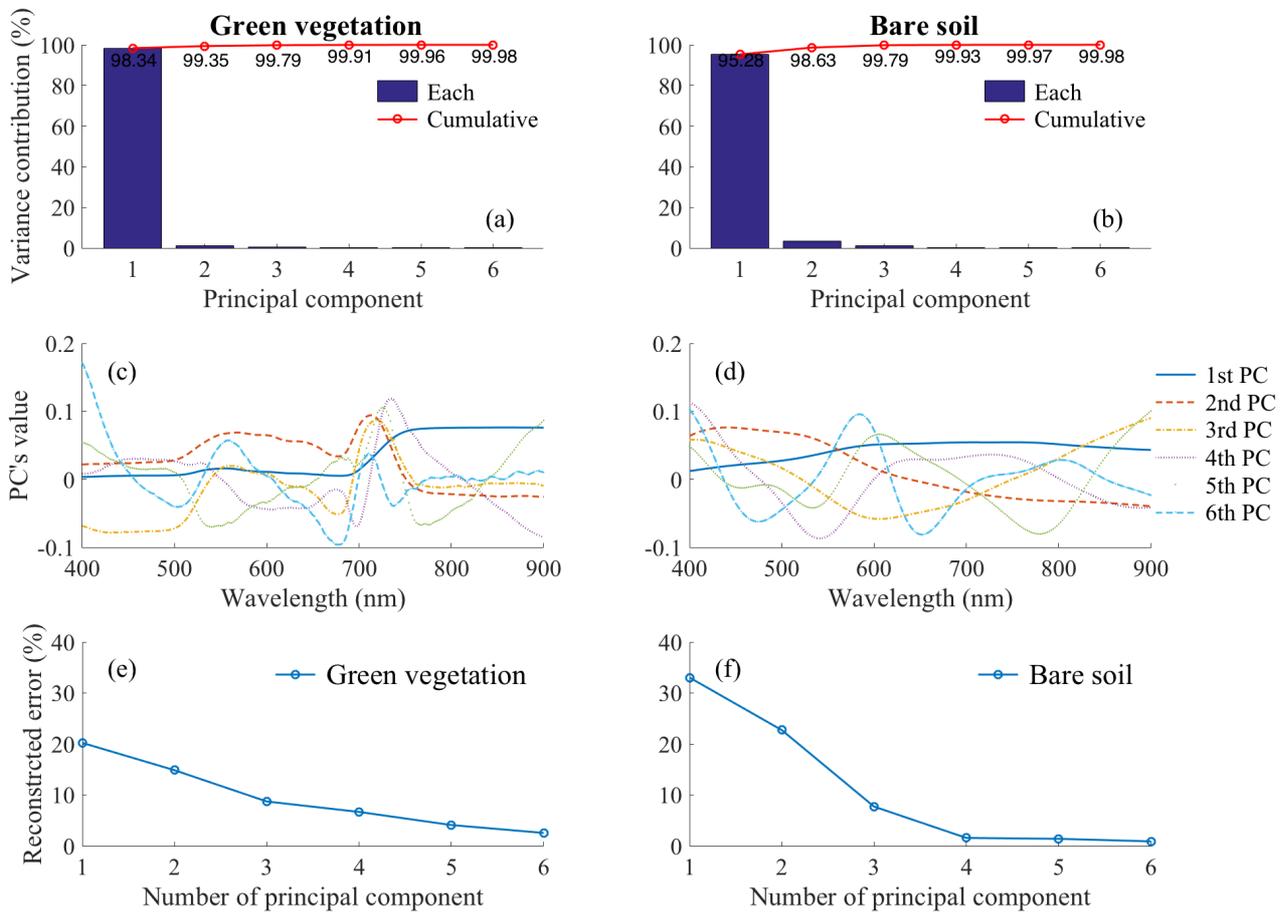

Figure 2. Principal component analysis results respectively for green vegetation dataset (left panel) and bare soil dataset (right panel). (a–b) Contribution of the first 6 principal components to the total variance; (c–d) spectra of the first 6 principal components; (e–f) the mean relative error in reconstructed reflectance of green vegetation and bare soil as a function of the number of principal components.

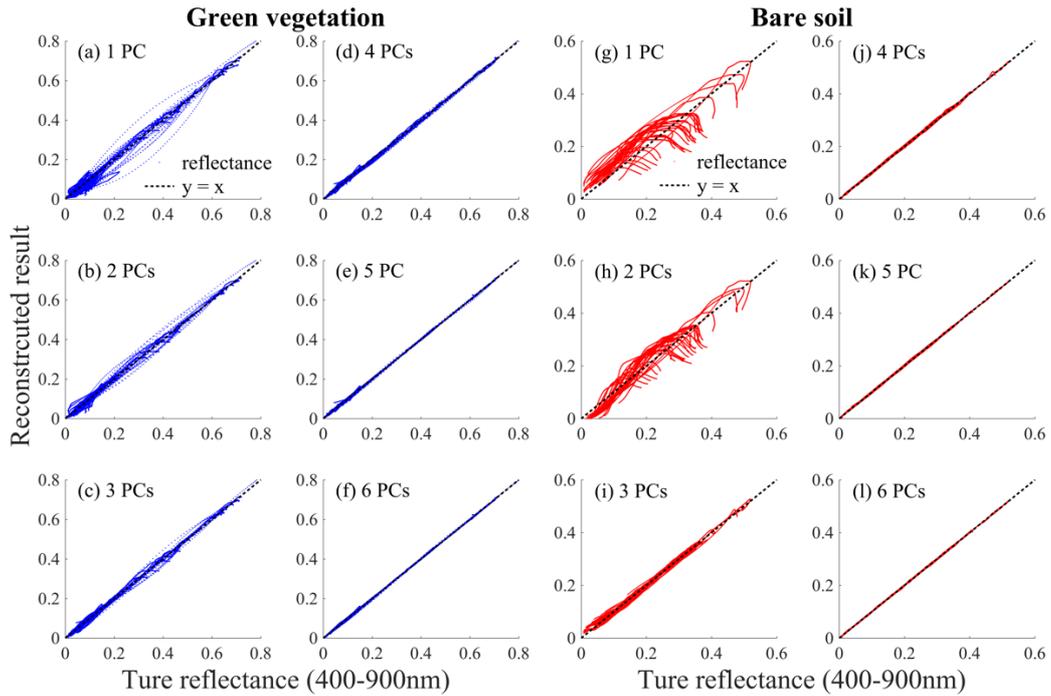

Figure 3. Panel (a–f): Scatterplots of the reconstructed reflectance with 1 PC to 6 PCs versus the true reflectance of green vegetation surface dataset over 501 spectral wavelengths in the range of 400–900nm. Panel (g–l): Same as panel (a)–(f), but for the bare soil surface dataset.

## 4.2 Spectral reconstruction results with selected bands

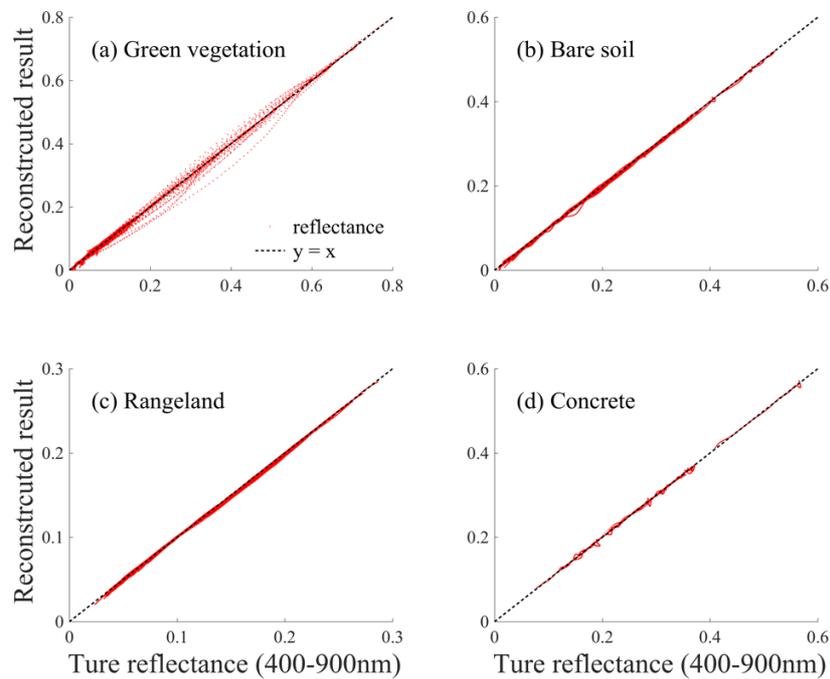

Figure 4. Scatterplots of the reconstructed reflectance with 6–7 wavelength bands versus the true reflectance of 4 different surface datasets over 501 spectral wavelengths in the range of 400–900nm. (a) Green vegetation, (b) Bare soil, (c) Rangeland, (d) Concrete.

Based on the extracted PCs from the spectral dataset in the range of 400–900 nm, we investigate the whole spectral reconstruction only employing the surface reflectance information of 6–7 typical wavelength bands, most of which are commonly used in satellite remote sensing. Figure 4 sequentially shows the scatterplots of the reconstructed reflectance based on the selected bands versus the true reflectance over 501 spectral wavelengths in the range of 400–900 nm for different datasets, and the correlation coefficients $R^2$ are also larger than 0.99. The selected wavelength bands and the mean reconstruction errors are listed in Table 1, in which the reconstruction error of green vegetation is about 3.3%, and the errors of other datasets are all smaller than 1.6%. In addition, Figure 5 illustrates the comparison of the averaged reconstructed spectra with the averaged true reflectance spectra in the spectral range of 400–900 nm, as well as the wavelength band used for spectral reconstruction. As can be seen from Figure 5, the reconstructed spectra are in a good agreement with the true spectra for different surface dataset.

Table 1. Wavelength band used for spectral reconstruction in the spectral range of 400–900 nm.

| **Spectral dataset** | **Wavelength band used** | **Reconstructed error** |
|---|---|---|
| Green vegetation | 440, 490, 555, 670, 760, 810, 865 nm | 3.3% |
| Bare soil | 440, 490, 555, 670, 760, 865 nm | 1.5% |
| Rangeland | 440, 490, 555, 670, 700, 810, 865 nm | 1.0% |
| Concrete | 400, 440, 490, 555, 670, 865 nm | 1.1% |

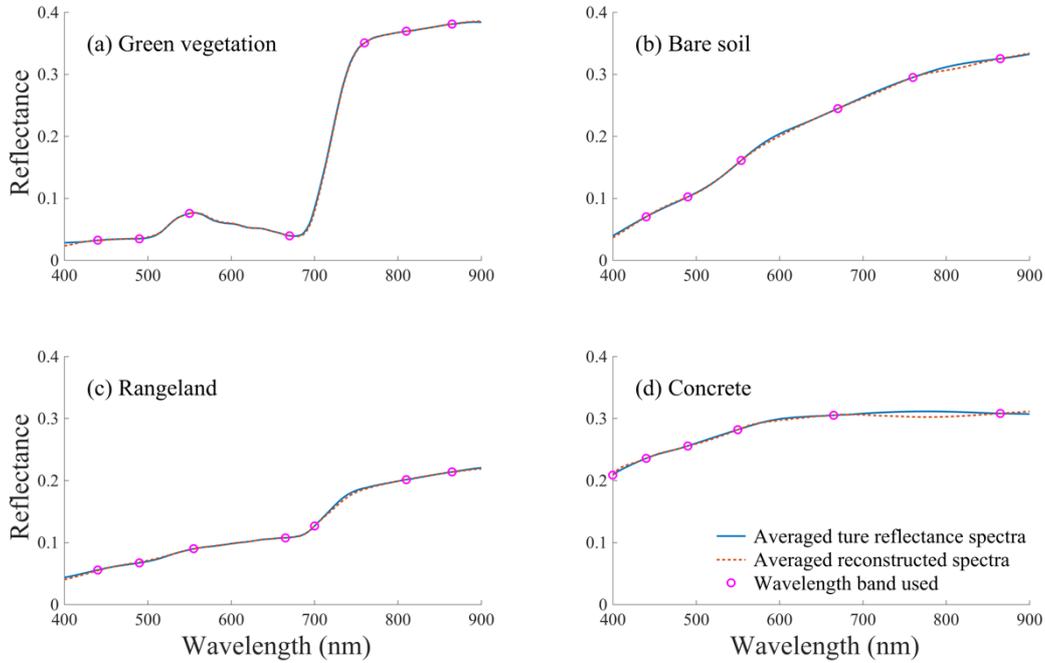

Figure 5. Comparison of the averaged reconstructed spectra with the averaged true reflectance spectra in the spectral range of 400–900 nm, as well as the wavelength band used for spectral reconstruction.

### 4.3 Spectral reconstruction results with linear combination

In order to test the spectral reconstruction based on the linear combination model, the surface information of four common bands for satellite remote sensing, including 490, 555, 670 and 865 nm[16-21], are used to the spectral reconstruction. Corresponding to Eq. (13), the reconstructed surface reflectance $\rho_b$ can be expressed as

$$\rho_b = a_1\rho_{490nm} + a_2\rho_{555nm} + a_3\rho_{670nm} + a_4\rho_{865nm} \tag{17}$$

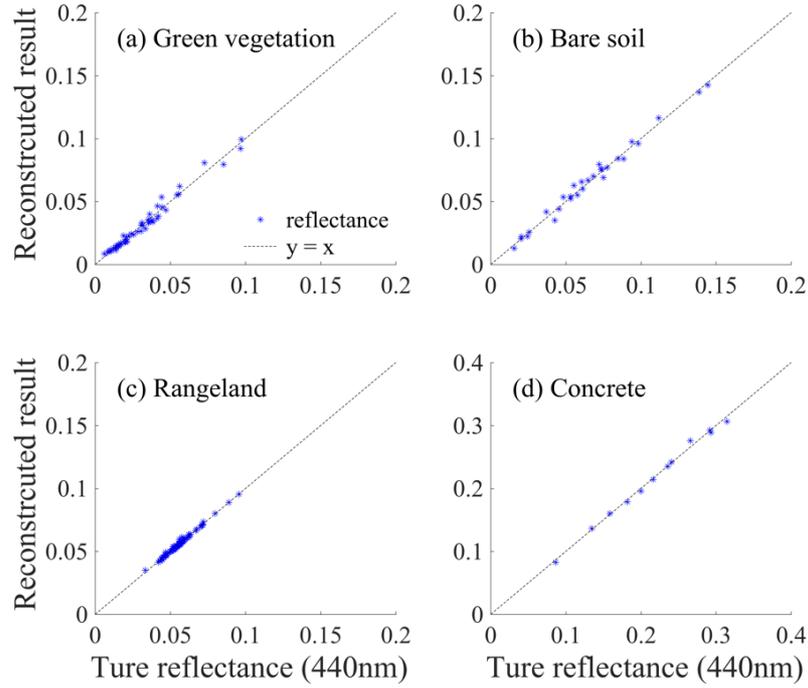

Figure 6. Scatterplots of the reconstructed reflectance at 440nm by the linear combination model with reflectance at 490, 555, 670, 865nm.

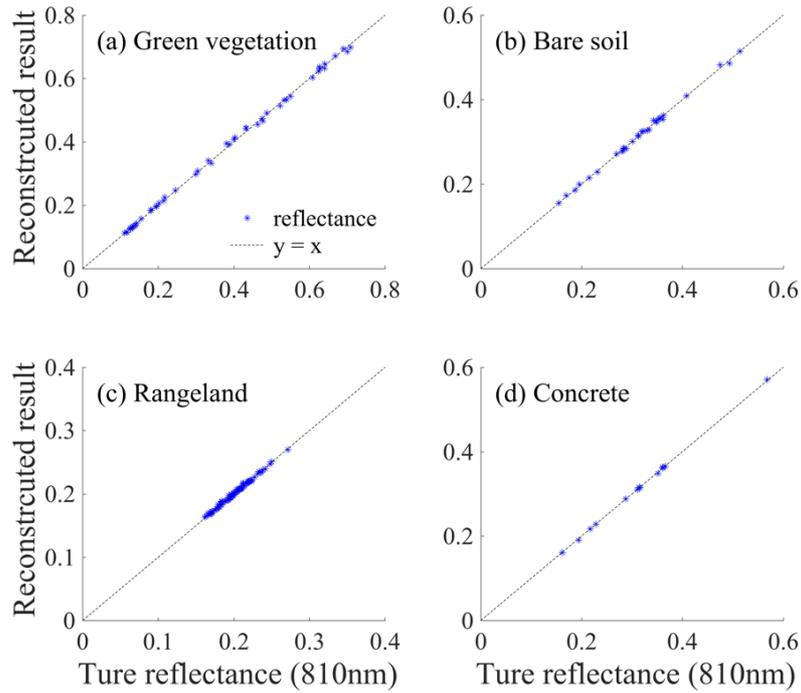

Figure 7. Same as the figure 6, but for the reconstructed reflectance at 810 nm.

By using the prior knowledge of spectra dataset and Eq. (14), the weighting coefficient vector $\boldsymbol{a} = [a_1, a_2, a_3, a_4]^T$ could be obtained. Taking the visible and near-infrared wavelength bands as an example, Figure 6 shows the scatter plots of the reconstructed and true surface reflectance at 440 nm, and Figure 7 shows the reconstructed reflectance results at 810nm. Correspondingly, Table 2 lists the weighting coefficients, reconstruction errors and correlation coefficients for these four different surface types. For the green vegetation, because the surface reflectance value at 440 nm is generally very small, although the relative error of spectral reconstruction is only 0.0024, the relative error is still 8.42%. However, while for the 810 nm wavelength, the absolute error of 0.0048 just corresponds to the relative error of 1.33%. Therefore, the relative error of the lower reflectance reconstruction is less than 8.5%, while the relative error of the higher reflectance reconstruction can be less than 1.5%, and the correlation coefficients ($R^2$) are all larger than 0.99.

Table 2. Take the reconstruction reflectance at 440 nm and 810 nm as an example, the weighting coefficients, the reconstructed errors and correlation coefficient for 4 different datasets respectively by the linear combination model with reflectance at 490, 555, 670 and 865 nm.

| Band | Surface type | $a_1$ | $a_2$ | $a_3$ | $a_4$ | Absolute error | Relative error | $R^2$ |
|---|---|---|---|---|---|---|---|---|
| 440 nm | Vegetation | 1.0664 | -0.0114 | -0.0879 | -0.0020 | 0.0024 | 8.42 % | 0.9902 |
| | Bare soil | 1.5545 | -0.6978 | 0.2261 | -0.0940 | 0.0030 | 5.40 % | 0.9961 |
| | Rangeland | 1.1660 | -0.1298 | -0.1115 | 0.0051 | 0.0007 | 1.39 % | 0.9950 |
| | Concrete | 1.4290 | -0.2016 | -0.5029 | 0.2619 | 0.0036 | 1.64 % | 0.9988 |
| 810 nm | Vegetation | 0.6868 | 0.2354 | -0.8808 | 0.9528 | 0.0048 | 1.33 % | 0.9996 |
| | Bare soil | -0.0141 | -0.0368 | 0.3709 | 0.7110 | 0.0030 | 0.96 % | 0.9990 |
| | Rangeland | -0.3077 | 0.4684 | -0.1000 | 0.8921 | 0.0015 | 0.78 % | 0.9966 |
| | Concrete | 0.0403 | -0.0015 | 0.1152 | 0.8610 | 0.0014 | 0.52 % | 0.9999 |

## 5. CONCLUSIONS

Based on the USGS and ASTER spectral library, considering four spectral datasets including green vegetation, bare soil, rangeland and concrete, the principal component analysis (PCA) and spectral linear combination method are investigated to reconstruct the typical surface reflectance spectra in the spectral range of 400–900 nm. Meanwhile, the leave-one-out cross validation method is used to analyze and evaluate the spectral reconstruction results. This study can provide a reference value for the surface reflectance processing and spectral reconstruction in satellite remote sensing research, and the specific conclusions are as follows:

(1) For the four typical surface reflectance datasets in the spectral range of 400–900 nm, the variance contribution of the first six principal components can be up to 99.98%, and the contribution rate of the first principal components has been over 95%. If all of the wavelength information is employed, the first six principal components can meet the requirements of spectral reconstruction by PCA. For the green vegetation dataset, the spectral reconstruction error is less than 2%; for other three datasets, the spectral reconstruction errors are all less than 1%.

(2) Base on the principal components obtained from the PCA of spectral datasets, the spectral reflectance of all wavelength in the range of 400–900 nm can be realized by using the surface reflectance of 6–7 typical bands. The mean reconstruction error of green vegetation is about 3.3%, and the mean error of other three dataset are all less than 1.6%, which can effectively meet the needs of spectral reconstruction.

(3) Based on the spectral linear combination model, the surface information of four common used wavelength bands for satellite remote sensing, including 490, 555, 670 and 865 nm, are considered to reconstruct the reflectance of other bands. The reconstruction error of is less than 8.5% for the dark surface reflectance case, while the reconstruction error of is less than 1.5% for the bright case, and the correlation coefficient are all larger than 0.99, which can basically meet the needs of spectral reconstruction.

## ACKNOWLEDGEMENTS

This study was supported by the Open Fund of State Key Laboratory of Remote Sensing Science (Grant No. OFSLRSS201710), the STS Project of Chinese Academy of Sciences (Grant No. KFJ-STS-QYZD-022), Instrument Developing Project of the Chinese Academy of Sciences (Grant No. YZ201664), and the National Natural Science Foundation of China (Grant Nos. 41505022, 41871269).